\documentclass[aps,prl,twocolumn,a4paper,showpacs,groupedaddress]{revtex4}

\pdfoutput=1

\usepackage{multibbl}
\usepackage[pdftex]{graphicx}
\usepackage{amsmath,SIunits}
\usepackage[latin1]{inputenc}
\usepackage{latexsym,stmaryrd}
\usepackage[sort&compress]{natbib}
\bibpunct{}{}{,}{s}{}{,}

\makeatletter
\newcommand*{\citenst}[2][]{%
  \begingroup
  \let\NAT@mbox=\mbox
  \let\@cite\NAT@citenum
  \let\NAT@space\NAT@spacechar
  \let\NAT@super@kern\relax
  \renewcommand\NAT@open{[}%
  \renewcommand\NAT@close{]}%
  \citep{#2}%
  \endgroup
}
\makeatother

\begin{document}

\title{Nonuniversal intensity correlations in 2D Anderson localizing random medium}

\author{Pedro David Garc\'{i}a}
\email{garcia@nbi.ku.dk}
\author{Søren Stobbe}
\author{Immo Söllner}
\author{Peter Lodahl}
\email{lodahl@nbi.ku.dk}
\homepage{http://www.quantum-photonics.dk/}

\affiliation{Niels Bohr Institute,\ University of Copenhagen,\ Blegdamsvej 17,\ DK-2100 Copenhagen,\ Denmark}

\date{\today}

\small

\begin{abstract}

Complex dielectric media often appear opaque because light
traveling through them is scattered multiple times.\ Although the light scattering is a random process, different paths through the medium can be correlated encoding information
about the medium.\ Here, we present spectroscopic measurements of nonuniversal intensity
correlations that emerge when embedding quantum emitters inside a disordered photonic crystal that is found to Anderson-localize light.\ The emitters probe
in-situ the microscopic details of the medium, and imprint such
near-field properties onto the far-field correlations.\ Our findings provide new ways of enhancing
light-matter interaction for quantum electrodynamics and energy
harvesting, and may find applications in subwavelength diffuse-wave spectroscopy for biophotonics.

\end{abstract}


\maketitle

Correlations are of paramount importance in science and technology and offer fundamental insight into disparate disciplines ranging from establishing the role of inflation in the early universe \cite{CMB}, through mesocopic transport of electrons and photons in complex media \cite{Akkermans}, to the non-locality of quantum mechanics observed at the microscopic scale \cite{Aspect}.\ In photonics, recording correlation functions provides a powerful method of revealing information of the quantum state of light \cite{Loudon} and correlations may be exploited for imaging beyond the classical diffraction limit \cite{Pittman}.\ The emerging field of nanophotonics promises a number of new applications for controlling and manipulating
light.\ However, unavoidable inhomogeneities lead to random multiple scattering that requires a statistical description.\ Surprisingly, such multiple scattering can give rise to pronounced correlations between different propagation paths through the complex medium, at variance with the intuitive perception that random scattering fully scrambles
all information.\ Such correlations can be either of classical \cite{Rossum} or quantum \cite{quantum} character and are traditionally universal, i.e., they depend solely on a single parameter, the universal conductance, and are independent of the microscopic details of the medium.\ Placing a light source inside the medium creates a fundamentally new situation.\ In this case, nonuniversal correlations in the far field have been predicted \cite{Saphiro} that depend sensitively on the local environment of the embedded emitter \cite{Tiggelen}, and thus can be employed for
probing the local properties of a complex medium.\ Consequently, recording the nonuniversal correlation function has been proposed as a novel method for in-situ spectroscopy with
ultra-high resolution in a complex and disordered medium \cite{skipetrov}, and algorithms have been proposed for imaging a source in an opaque medium from measurements of the local electromagnetic field density \cite{Carminati_source}.\ In this Letter, we measure for the first time the nonuniversal correlations generated by single dipole emitters in a strongly scattering random medium.

\begin{figure}[b!]
  \includegraphics[width=8.2cm]{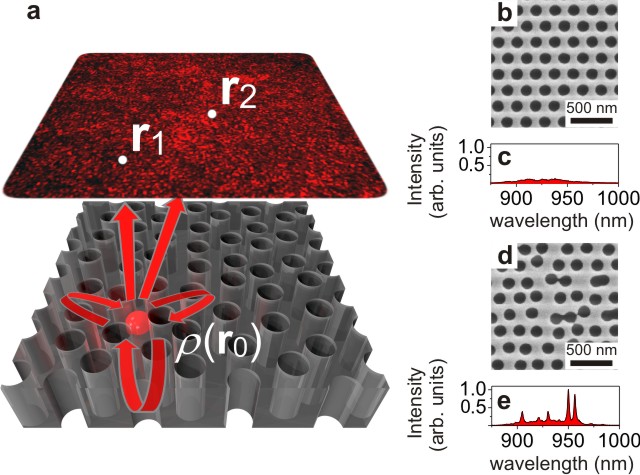}
      \caption{ \label{1} (Color online) \textit{Intensity fluctuations in disordered photonic crystals.} (a), Illustration of a far-field intensity speckle pattern originating from a single emitter in a membrane perforated by a disordered arrangement of holes.\ The red arrows illustrate the link between the near-field environment of the emitter, the local density of optical states $\rho(\textbf{r}_0)$, and the far-field intensity pattern.\ (b), (d), Scanning electron micrographs (top view) of photonic crystal membranes with introduced disorder on the hole positions of $\delta = 0\%$ and $\delta = 12\%$, respectively.\ (c), (e), QD photoluminescence spectra collected on the two samples.\ The spectra are scaled equally for better comparison.\ The bright peaks in the spectra are signatures of Anderson-localized modes.}
    \end{figure}

\begin{figure*}[t]
  \includegraphics[width=17cm]{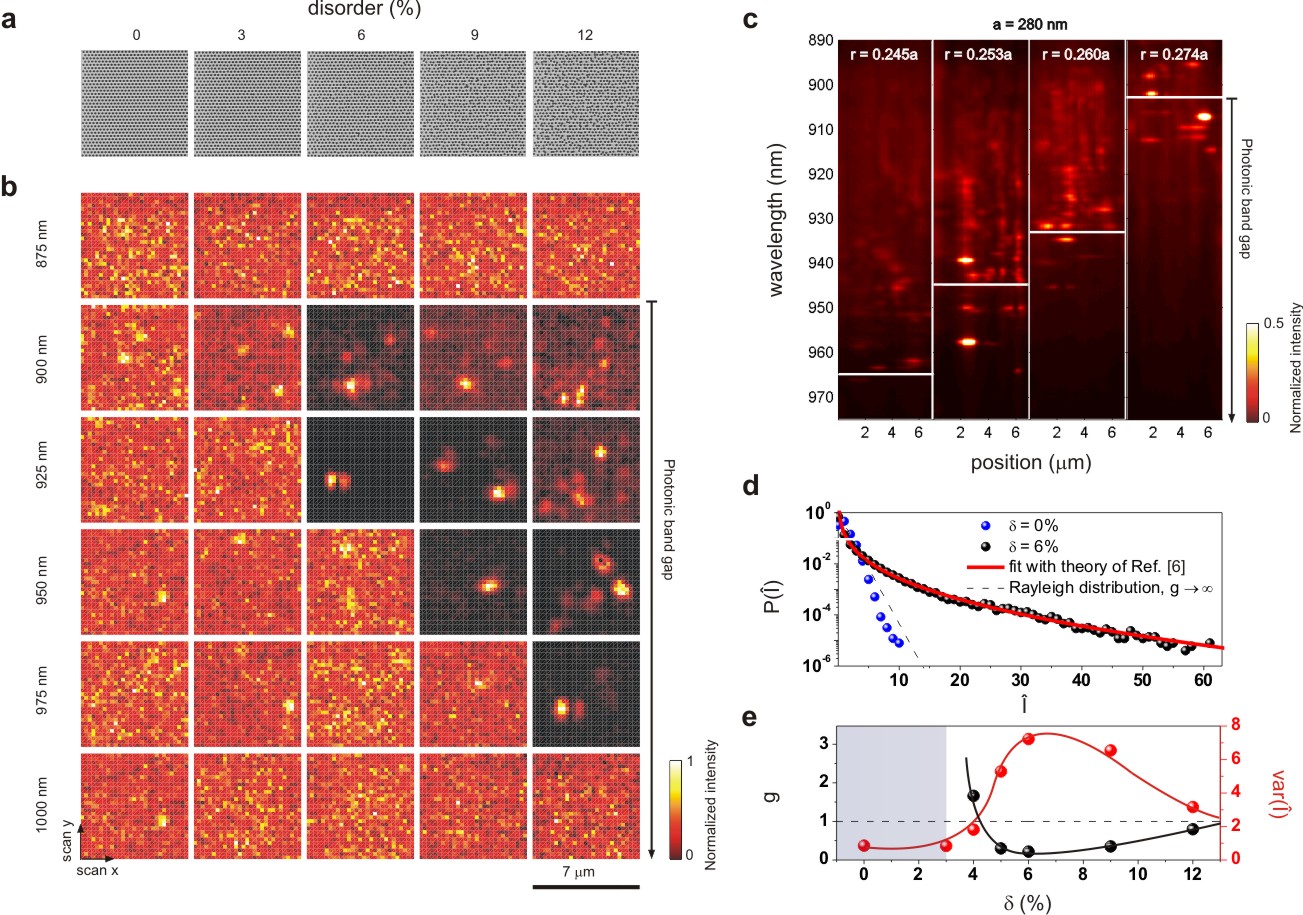}
    \caption{ \label{2} (Color online) \textit{Spatial and spectral mapping of the far-field intensity pattern.} (a), Scanning electron micrographs (top view) of photonic crystal membranes with $a=280\,\text{nm}$, $r=0.274a$, and with the degree of disorder varying between $\delta = 0\%$ and $\delta = 12\%$.\ (b), Normalized high-power photoluminescence spectra collected while scanning the samples with a wavelength bin size of $1\,\text{nm}$.\ (c), High-power photoluminescence spectra collected while scanning photonic crystals with $\delta = 6\%$, $a = 280\,\text{nm}$, and different hole radii.\ The horizontal lines represent the spectral positions of the high-energy band edge of the unperturbed structures.\ (d), Normalized intensity distribution in a photonic crystal with $\delta = 0\%$ and $\delta = 6\%$.\ The large deviation from Rayleigh statistics (dashed line) in the disordered case is due to Anderson localization.\ The solid red line represents the best fit to the theory of Ref.~\citenst{Rossum}.\ (e), Dimensionless conductance, $\textit{g}$, and variance of the intensity fluctuations for various amounts of disorder.\ The lines are guides to the eye and the shaded area represents the samples where fully developed speckles were not observed for the sample size of $7\,\micro\text{m} \times 7\,\micro\text{m}$ employed in the present experiment. Anderson localization is observed for $\text{var}(\widehat{I}) \geq 7/3$ (dotted line).}
    \end{figure*}

Figure~\ref{1}(a) illustrates an intensity speckle pattern generated by a dipole source embedded in a disordered photonic crystal membrane where random noise in the position of the air holes gives rise to strong multiple scattering.\ The generated correlations are gauged by the intensity correlation function (denoted $C_0$) that connects the intensities at two different spatial positions \cite{Saphiro,Tiggelen,Carminati} $\textbf{r}_1$ and $\textbf{r}_2$,
\begin{equation}\label{correlation}
C_0 \cong \frac{ \langle
 I(\textbf{r}_1)I(\textbf{r}_2) \rangle_{\text{ff}}
 - \langle I(\textbf{r}_1)\rangle_{\text{ff}} \langle
I(\textbf{r}_2)\rangle_{\text{ff}}}{\langle I(\textbf{r}_1)\rangle_{\text{ff}} \langle
I(\textbf{r}_2)\rangle_{\text{ff}}} = \frac{\text{Var}[\rho(\textbf{r}_0)]_{\text{nf}}} {{\langle \rho(\textbf{r}_0) \rangle}^2 _{\text{nf}}},
\end{equation}
where the brackets $\left< \ldots \right>_{\text{nf}/\text{ff}}$ denote spatial averaging in the near- and far-field, respectively.\ Eq.~(\ref{correlation}) holds after spectral averaging where all contributions from standard (i.e. universal) mesoscopic correlations \cite{Sheng} vanish, which is the case in the present experiment.\ The $C_0$ correlation has infinite range, i.e., it is independent of the observation points $\textbf{r}_1$ and $\textbf{r}_2$ in the far field.\ Eq.~(\ref{correlation}) demonstrates that $C_0$ is linked to local near-field properties of the random medium through the local density of optical states $\rho(\textbf{r}_0)$ at the position $\textbf{r}_0$.\ This provides a way of extracting $C_0$ for light since in optics, as opposed to, e.g., electronics, excellent point sources exist such as single atoms or quantum dots (QDs), and their spontaneous-emission dynamics probes $\rho(\textbf{r}_0)$.\ $C_0$ correlations are distinct from other multiple scattering correlations due to the dependence on the local properties and this nonuniversality has been proposed as a method for subwavelength imaging of a random medium \cite{skipetrov}.

\begin{figure*}[t]
  \includegraphics[width=17cm]{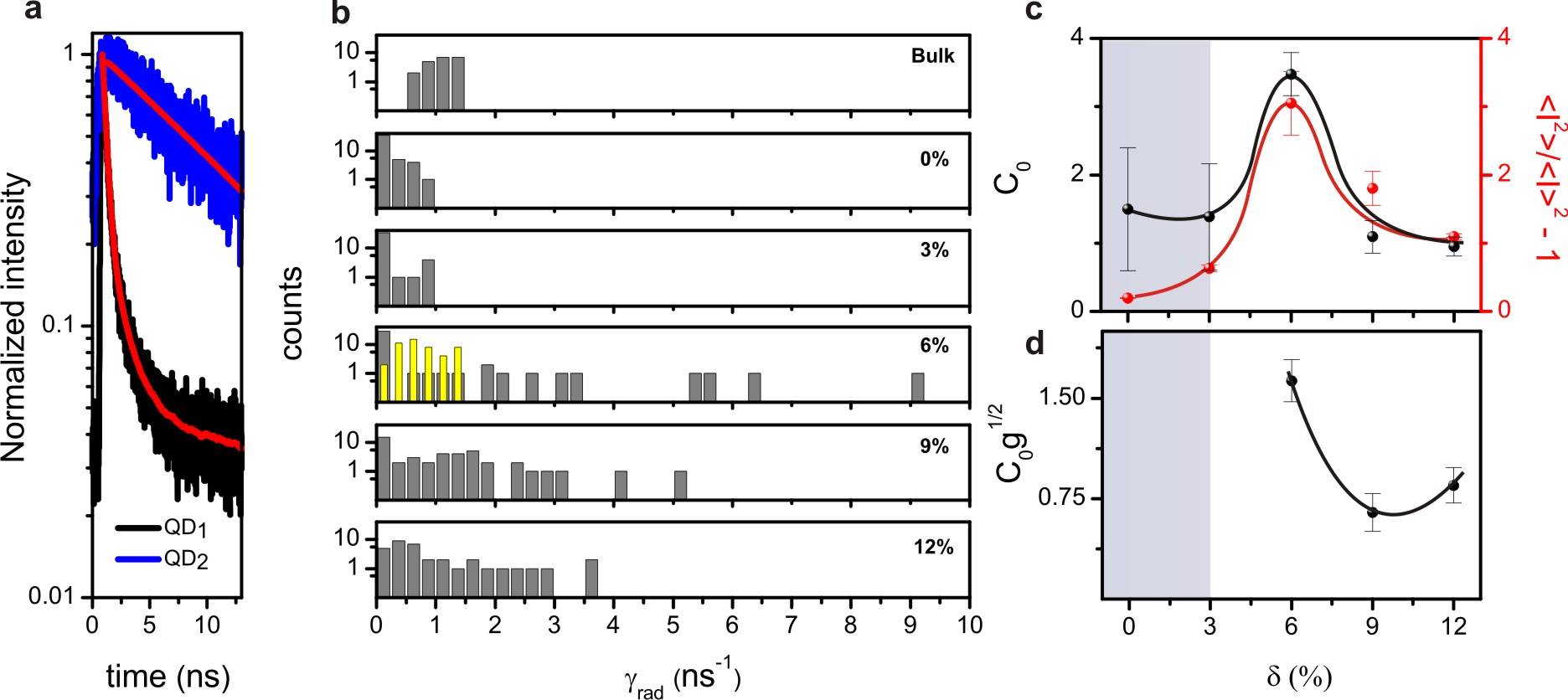}
    \caption{ \label{3} (Color online) \textit{Exploring the near-field by probing dynamic properties of single emitters.} (a), Decay curves for two QDs at different positions in a photonic crystal with $\delta = 6\%$ and the fits to bi-exponential models (red lines).\ $\text{QD}_1$ is enhanced by the coupling to an Anderson-localized mode while $\text{QD}_2$ is inhibited.\ (b), Radiative decay rate ($\gamma_\text{rad}$) distributions measured at $\lambda = 910\pm3\,\text{nm}$ in the bulk membrane and in photonic crystals with $a=280\,\text{nm}$ and $r=0.274a$ affected by different amounts of disorder.\ The yellow bars show the $\gamma_\text{rad}$ distribution measured in a photonic crystal with the same lattice constant but with $r=0.245a$.\ (c), $C_0$ correlation function as extracted from the $\gamma_\text{rad}$ distributions (black symbols) and from the spatial intensity fluctuations (red symbols).\ The shaded area represents the region of not fully developed multiple scattering and the lines are guides to the eye.\ (d), Variation of $C_0 \times \sqrt{g}$ with degree of disorder illustrating the nonuniversality of  $C_0$. The curve is a guide to the eye.}
    \end{figure*}

Probing the spatial fluctuations of $\rho(\textbf{r}_0)$ with single emitters constitutes an efficient way of extracting $C_0$.\ Founding experiments on modified emission dynamics in random media employed ensembles of emitters \cite{Vos,CarminatiPlasmons,Ricca}, which do not enable extracting $\rho(\textbf{r}_0)$ in a complex medium and therefore the $C_0$ correlation could not be obtained.\ Experiments on single emitters have so far concentrated on cavity quantum-electrodynamics studies in one-dimensional disordered photonic crystal waveguides \cite{Luca}, where the collection of statistics required for obtaining the $C_0$ correlation could not be obtained.\ Here, we present measurements of the $C_0$ correlation in disordered photonic crystal membranes in the regime of Anderson localization \cite{Anderson}.\ We employ single InAs QDs that have recently been shown to be reliable probes of $\rho(\textbf{r}_0)$ in dielectric nanostructures when the internal fine-structure splitting and the non-radiative processes of the QD are properly accounted for \cite{Qin}.\ We observe the transition to the regime of two-dimensional Anderson localization by varying the degree of disorder of our samples, which gives rise to a pronounced $C_0$ correlation.\ Deep in the localization regime the $C_0$ correlation is found to be very strong, reaching a value as high as $3.5$, which is several times larger than is expected for scalar waves in the most strongly scattering 3D structures reported to date \cite{Saphiro,GaP,AL_maret}.\ Finally, we demonstrate that the statistical properties of the Anderson-localized modes, and therefore the $C_0$ correlation, can be controlled and fine tuned by employing the underlying dispersion of the photonic crystal.\ The aid of a photonic band edge for Anderson localization of light was the original motivation of S. John for proposing photonic band gap crystals \cite{John}, and our experiment provides to our knowledge the first direct experimental demonstration of this effect.

To probe the far-field properties of our samples, we excite many QDs embedded in the system and measure the far-field emission pattern.\ Fig.~\ref{1}(b) and \ref{1}(d) show top-view scanning electron micrographs of photonic crystals where the holes have been randomly displaced corresponding to $\delta = 0\%$ and $\delta = 12\%$, where $\delta$ is the standard deviation of the hole positions with respect to the lattice constant.\ Fig.~\ref{1}(c) and \ref{1}(e) show recorded photoluminescence intensity spectra of the light emitted from the QDs under strong pump power conditions where the photonic modes of the sample are revealed, as opposed to the single QD emission lines seen at low excitation power \cite{Smolka}.\ A smooth emission spectrum is observed for no intentional disorder $(\delta=0\%)$, which resembles the inhomogeneously broadened emission spectrum of the QDs.\ Increasing the degree of disorder leads to larger intensity fluctuations and for $\delta \ge 5\%$ pronounced resonances are observed.\ This is due to the imperfections in the hole position that lead to random multiple scattering eventually inducing Anderson localization if the disorder is sufficiently pronounced.\ Figure~\ref{2}(a) and \ref{2}(b) show a range of photonic crystals with various degrees of disorder and the far-field intensity collected while raster scanning the samples.\ By varying either the hole radius (Fig.~\ref{2}(c)) or the lattice constant (not shown) for a fixed amount of disorder $(\delta = 6\%)$, we observe that the spectral region where the cavities appear follows the photonic crystal band edges, which in turn implies that the $C_0$ correlations can be tuned by taking advantage of the underlying dispersion, as will be discussed later.\ The Anderson-localized cavities appear in a spectral range close to the band edge \cite{John} $(\lambda = 903\,\text{nm})$.\ This range broadens with the amount of disorder, which is a signature of the so-called Lifshitz tail that was originally proposed for electrons \cite{Lifshitz} and originates from the broadening of the band edge with disorder \cite{Huisman}.

The signatures of the nonuniversal $C_0$ can be demonstrated by extracting from the far-field intensity measurements the universal dimensionless conductance $\textit{g}$.\ $\textit{g}$ is the universal scaling parameter that governs the statistical aspects of light transport in a random medium \cite{Chabanov}, and the threshold for localization in absence of loss is \cite{Sheng} $\textit{g} = 1$.\ $\textit{g}$ can be obtained from the probability distribution describing the intensity fluctuations \cite{Rossum} $P(\widehat{I} \equiv I/\langle I
\rangle)$, which is extracted experimentally from the intensity scans shown in Fig.~\ref{2}(b), and displayed in Fig.~\ref{2}(d). See Supplementary Information for further details.\ In the highly disordered case, the distribution significantly deviates from the Rayleigh statistics expected for diffusive propagation.\ Excellent agreement with the theory of Ref.~\citenst{Rossum} is observed over five orders of magnitude, which is quite remarkable since the theory is formally derived only in the mesoscopic regime at the onset of Anderson localization.\ Here we find it valid also deeply inside the localized regime in agreement with previous experimental work on ultrasound \cite{Sound}.\ From the comparison with our data we obtain $\textit{g} = 0.22 \pm 0.07$ for $\delta = 6\%$.\ We note that pronounced loss may reduce the value of $\textit{g}$, but cannot explain a large variance.\ In the presence of loss, $\text{var}(\widehat{I}) \geq 7/3$ serves as a localization criterion for a multimode waveguide \cite{Chabanov} that can be extracted directly from the experimental data without assuming any model.\ Note that for this criterion it is assumed that the intensity fluctuations of the light leaking vertically out of the structure obey the same statistics as the transmission.\ Numerical modeling of 1D systems has proven the validity of such an assumption, and the required variance of intensity fluctuations for Anderson localization was in fact found to be smaller than the transmission fluctuations \cite{Smolka_PhD}.\ The Anderson-localization criterion is clearly fulfilled in our experiments, and we observe, e.g., $\text{var}(\widehat{I}) = 7.2$ for $\delta = 6\%.$\ Figure~\ref{2}(e) displays $\textit{g}$ and $\text{var}(\widehat{I})$ versus the degree of disorder and shows that an optimum degree of disorder exists $(\delta = 6\%)$ where $\textit{g}$ is smallest and $\text{var}(\widehat{I})$ largest, thus light localization strongest.\  Such an optimum has been predicted theoretically for 3D disordered photonic crystals \cite{Conti}, and reflects the fact that disorder is introduced as a perturbation to an underlying ordered structure that aids Anderson localization.\ For large degrees of disorder, the 2D band gap closes and band edge localization disappears.

The near-field environment is investigated by probing dynamic properties of the embedded emitters.\ For weak light-matter interaction strengths, the radiative decay rate of an emitter is directly proportional to the projected local density of states at the position $\textbf{r}_0$ of the emitter, i.e., $\gamma_\text{rad}(\textbf{r}_0,\lambda)\propto|\textbf{p}(\lambda)|^2 \rho(\textbf{r}_0,\lambda)$, where $\textbf{p}(\lambda)$ is the transition dipole moment, and $\lambda$ is the wavelength of the optical transition.\ Thus the QDs embedded in the samples are direct probes of $\rho(\textbf{r}_0)$ and from its spatial fluctuations the $C_0$ correlation is extracted.\ To that end, we measure single QD emission decay curves by time-resolved spectroscopy, as shown in Fig.~\ref{3}(a), and perform the ensemble average by repeating the measurements for QDs placed at different positions.\ The measured decay curves are biexponential revealing the fine structure of the lowest exciton state \cite{Bayer}.\ It splits into bright and dark exciton states, which differ in their total angular momentum being 1$\hbar$ or 2$\hbar$, respectively.\ For this reason, a dark exciton cannot recombine through an optical dipole transition unless it becomes bright after a spin-flip process \cite{Smith}.\ From the biexponential model we extract the spin-flip rate between the dark and bright excitons, the non-radiative, and the radiative decay rate from the bright exciton of different QDs, as explained in detail in Ref.~\citenst{Qin}.\ According to our measurements, only the radiative decay rate is affected by disorder while the other decay rates remain unperturbed (not shown).

Figure~\ref{3}(b) displays the histogram of $\gamma_\text{rad}$ measured in an unpatterned substrate (bulk) and in photonic crystals with different amounts of disorder.\ All the decay curves are measured in a narrow wavelength range ($\lambda = 910\pm3\,\text{nm}$) in order to avoid any effects of the wavelength dependence of the transition dipole moment \cite{jeppe_energy}.\ For no intentional disorder ($\delta = 0\%$), we observe a strong suppression of the spontaneous emission rate compared to bulk due to the 2D photonic band gap.\ With increasing amount of disorder, the distribution of decay rates broadens tremendously, cf.~Fig.~\ref{3}(b), and largely enhanced rates are observed in the regime of Anderson localization $(\delta \geq 6\%)$.\ For $\delta = 6\%$ in particular, we observe a maximum rate of $\gamma_{\text{rad}}= (9.13 \pm 0.02)\,\text{ns}^{-1}$ and a minimum rate $\gamma_{\text{rad}}=(0.08 \pm 0.06)\,\text{ns}^{-1}$, see Fig~\ref{3}(a), which corresponds to a Purcell enhancement as high as $114$.\ To our knowledge this is the largest Purcell enhancement ever reported in any nanophotonic quantum electrodynamics experiment, and illustrates the great potential of disordered photonic materials for light-matter enhancement.\

Figure~\ref{3}(c) compares the measured $C_0$ correlation function to the intensity fluctuations of the Anderson-localized modes as recorded from the far-field measurements shown in Fig.~\ref{2} after averaging over the same wavelength range as we measured $\gamma_\text{rad}$.\ A close similarity is observed, which is surprising since the intensity-fluctuation measurements probe the localized modes after excitation by many QDs, which is not a direct probe of $\rho(\textbf{r}_0)$ as required for $C_0$ measurements.\ Having extracted both $g$ and $C_0$ independently enables quantifying the nonuniversality of the latter.\ In the regime where scattering can be treated perturbatively $C_0$ is inversely proportional to $\sqrt{g}$, and deviation from this scaling is a signature of nonuniversality \cite{Saphiro,skipetrov}.  In Fig.~\ref{3}(d) we plot $\sqrt{g} \times C_0$, which clearly deviates from the universal scaling as a result of the breakdown of the perturbative scattering theory in the localized regime and the fact that $C_0$ depends on the near-field properties of the source.\ Consequently $C_0$ is sensitive to the near-field environment of the source, i.e., information about the correlation length of the local dielectric function could be extracted.\ This sensitivity may pave a route to subwavelength diffuse-wave microscopy of disordered opaque photonic media \cite{skipetrov}.\ Finally we have observed that the $C_0$ correlation function can be tuned by taking advantage of the underlying order of the photonic crystal lattice.\ In Fig.~\ref{3}(b) we show for $\delta = 6\%$ the decay rate distributions for two disordered photonic crystals with the same lattice constant but different hole radii.\ By changing the hole radii we can tune the photonic band edge to be out of resonance with the quantum dot luminescence spectrum (corresponding to yellow data of Fig.~\ref{3}(b)) thereby significantly reducing the strength of multiple scattering and hence $C_0$.\ We obtain $C_0=0.62(8)$, which is five times smaller than the value measured at the band edge, revealing that the $C_0$ correlations can be efficiently tuned by controlling the dispersion.

We have presented the experimental observation of nonuniversal correlations when embedding light emitters in a complex random medium.\ The ability to measure these correlations opens the possibility of probing locally the complex properties of random media.\ Such insight is essential for using random media for lasing \cite{Cao} or quantum optics experiments where potentially a single photon and a single quantum dot can become entangled by exploiting disorder for enhancing light-matter interaction \cite{Tyrrestrup}.\ The nonuniversality of the $C_0$ correlations may be exploited for superresolution spectroscopy inside an opaque medium, since it gains insight into the near-field properties.\ Random media are currently also considered as a way to harvest light for solar energy applications by employing enhanced local absorption \cite{solar_cell}, and our work and method may be applied for engineering the efficient absorption of light.

\section{Acknowledgements}

We gratefully acknowledge financial support from the Villum Foundation, The Danish Council for Independent
Research (Natural Sciences and Technology and Production Sciences), and the European Research Council (ERC consolidator grant "ALLQUANTUM").


\newpage

\section{Supplementary information}

\textbf{Fabrication of the photonic crystal structures.}

The samples are GaAs membranes containing a single layer of self-assembled InAs quantum dots (QDs) in the
center with density $80 \,\micro\text{m}^{-2}$.\ An ordered
triangular lattice of holes is etched in the structure creating a
2D photonic band gap that suppresses in-plane propagation of
light.\ We investigate a set of samples with dimensions
$7\,\micro\text{m} \times 7\,\micro\text{m}$, membrane height $h=(150 \pm 5)\,\text{nm}$, a range of lattice constants $220\,\text{nm}<\textit{a}<280\,\text{nm}$, and hole radii
$0.24\textit{a}<\textit{r}<0.3\textit{a}$.\ The hole positions are randomly
varied using a Gaussian random number generator
function.\ The amount of disorder $\delta$ is characterized by
the standard deviation of the hole position with respect to the
lattice constant varying from $\delta = 0\%$ to $\delta = 12\%$.\ The photonic band structure was calculated for $\delta = 0\%$ using the software MIT
PHOTONIC BANDS.\ The refractive index of GaAs, $n=3.55 - 3.44$, is calculated~\cite{bands} for $T=10\,\text{K}$ in
the spectral range $\lambda = 850\,\text{nm} - 1000\,\text{nm}$.\

\textbf{Photoluminescence experiments.}

For optical characterization we use a confocal micro-photoluminescence setup to excite QDs within a diffraction-limited region and collect the emitted light (details can be found in Ref.~\citenst{Toke}).\ The samples are cooled to a temperature of $T = 10\,\text{K}$ in
a helium flow cryostat and the sample position is controlled with
stages with a resolution of $0.3\,\micro\text{m}$.\ To efficiently
excite Anderson-localized modes, we employ the inhomogeneously
broadened spectrum from saturated QDs pumped at a high
excitation power density of $P = 1\text{kW}/\text{cm}^2$.\ Emission spectra are collected in a wide wavelength range $850\,\text{nm} - 1000\,\text{nm}$ with a spectrometer ($0.15\,\text{nm}$ resolution) equipped with a CCD camera while scanning the excitation and collection objective along the photonic crystals.\ The intensity probability distribution $P(\widehat{I})$ is measured by collecting the intensity $I^{\lambda}_{x,y}$ at each spatial position $(x,y)$ with a spatial binning size of $0.23
\,\micro\text{m}$ and at different wavelengths with a binning size of $1 \,\text{nm}$.\ Subsequently, an average over the wavelength range $850\,\text{nm} - 1000\,\text{nm}$ is performed.

The dimensionless conductance, $\textit{g}$, is obtained by fitting the experimental intensity probability distribution with the full speckle intensity distribution obtained in Ref.~\citenst{Rossum_S} in the regime of perturbative scattering and the absence of absorption as:
\begin{equation}\label{intensity_distributoion}
P(\widehat{I})= \int_{- i \infty}^{i \infty} \! \frac{\mathrm{d} x}{\pi i} K_0 (2 \sqrt{-\widehat{I}x}) e^{-\Phi_{con}(x)}
\end{equation}
where $K_0$ is a modified Bessel function of second kind and $\Phi_{con}(x)$ is obtained by assuming plane-wave incidence as:
\begin{equation}\label{intensity_distributoion}
\Phi_{con}(x)= gln^2(\sqrt{1 + \frac{x}{g}} +\sqrt{\frac{x}{g}})
\end{equation}

\textbf{Probing the radiative decay rate.}

The QD decay rate is measured by time-resolved photoluminescence spectroscopy at an excitation power density of $\text{P} = 24\text{W}/\text{cm}^2$, which is below the saturation level of a single QD.\ The QDs are excited with short optical pulses (2ps) at a $75\,\text{MHz}$ repetition rate.\ Single QD lines are detected by spectral filtering and their decay curves are recorded with an avalanche photo diode.\ By accounting for the
exciton fine structure, it is possible to extract the radiative
decay rate and therefore to separate effects from nonradiative
recombination and spin-flip processes.\ The decay curves are fitted with a biexponential
model $I(t)=A_\text{f}e^{-\gamma_{f}t}+A_\text{s}e^{-\gamma_{s}t}$,
where subscripts $\textit{s}$ and $\textit{f}$ denote slow and fast
decay components, respectively.\ From $\gamma_{f}$, $\gamma_{s}$
and $A_\text{f}/A_\text{s}$ we determine the spin-flip rate
between the dark and bright excitons, the non-radiative, and the
radiative decay rate of the bright excitons by employing the method detailed in Ref.~\citenst{Qin_S}.\ We note that the proper extraction of the radiative rate, as opposed to merely studying the fitted rates $\gamma_{f}$ or $\gamma_{s}$, is essential in order to probe the local density of optical states and thereby obtain the $C_0$ correlation function.

\end{document}